\documentclass[twocolumn,prl,showpacs,amsmath,amssymb]{revtex4}

\usepackage[dvipdfmx]{graphicx}
\usepackage{dcolumn}
\usepackage{bm,color}

\begin{document}

\preprint{APS/}

\title{Inertial energy dissipation in nuclear dynamics}

\author{Yoritaka Iwata$^{1}$}
\author{Takashi Nishikawa$^{2}$}
\email{iwata_phys@08.alumni.u-tokyo.ac.jp}

\affiliation{$^{1}$Faculty of Chemistry, Materials and Bioengineering, Kansai University, Osaka 564-8680, Japan}
\affiliation{$^{2}$Nuclear Engineering, Ltd., Osaka 550-0001, Japan}

\date{\today}

\begin{abstract}
We present the TDDFT+Langevin model that incorporates microscopic time-dependent density function theory (TDDFT) with macroscopic Langevin model.
By extracting the energy-dependent dissipation effect from the TDDFT dynamics, quantum effects are introduced to the Langevin-type stochastic fission analysis.
In this paper, by means of Skyrme nuclear effective interactions, the energy-dependent friction coefficients to be used in Langevin calculations are provided individually for 25 fission nuclei chosen from uranium, plutonium, curium, californium, and fermium isotopes.
The validity of the energy-dependent friction coefficients are confirmed by comparing to the existing experimental fission fragment yields.
In conclusion, depending on the energy, the transition of energy dissipation mechanism from conventional viscous energy dissipation to inertial energy dissipation is shown.
\end{abstract}


\maketitle
\section{Introduction}%
Nuclear fission plays an indispensable role in the nuclear engineering, while the microscopic and quantum understanding of fission dynamics is still an open problem (for the recent developments in the relevant microscopic approaches, see \cite{18scamps,18simenel,19stevenson,19sekizawa,20bender}). 
Energy dissipation contributes substantially to the fission dynamics, as the timescale of pre-scission neutron emission is similar to that of microscopic dissipation with its time scale $\sim$ 10$^{-20}$ to 10$^{-21}$s, where the time-scale of each reaction dynamics is an important ingredient of heavy-ion collisions including nuclear fission dynamics (see \cite{20simenel} for a recent microscopic work, and see \cite{13iwata-4} for a gross view).
Although larger dissipation is expected to enhance the fission probability, such a basic information has not been quantitatively confirmed by the microscopic reaction dynamics.
Furthermore the synthesis of chemical elements heavier than iron and that of superheavy nuclei are associated with the dissipative reaction dynamics. 
One of the most successful theoretical frameworks for dissipative nuclear fission dynamics is the Langevin model.
In the Langevin calculations, the dissipation effect has been introduced by adjusting several free parameters.
Phenomenological parameter fittings have been used to predict unknown phenomena based on the parameters well-reproducing the known experimental results.
However, phenomenological parameter fitting does not necessarily work to explore the unknown nuclei, and the non-empirical treatment is required instead.
Our purpose is to introduce less-phenomenological friction coefficients to four-dimensional Langevin calculations, which are obtained by microscopic time-dependent density functional theory (TDDFT). 
Based on the TDDFT + Langevin model including the fully-introduced one-body dissipation, the inertial dissipation effect is discovered in quasi-fission dynamics for the first time.

There are two types of energy dissipation: viscous dissipation and inertial dissipation.
The amplitude of force causing the viscous dissipation is proportional to the velocity, while that causing the inertial dissipation is proportional to the square of the velocity.
The inertial dissipation draws a special attention, as it is a higher order effect to be included in nonlinear dynamics appearing in the complex physical systems (for a recent study, see \cite{20mostert}).
Usually, inertial dissipation is observed in the high velocity situations, e.g., fluids with large Reynolds numbers.
In the present cases with colliding heavy ions (including only finite-body dynamics) in the femto-meter scale, we show that the inertial friction plays a role also in the slow velocity situations, and the viscous friction is located in the intermediate velocity situations. 
It is notable here that the low-energy heavy-ion collisions with those relative velocities less than 1/3 of speed of light are nonlinear processes, while intermediate or high energy heavy-ion collisions are equal to or similar to linear processes to be treated by non self-consistent scattering theory.
That is, the appearance of inertial dissipation does not essentially arise from the velocity of medium, but from the nonlinearity of the system. 

The usefulness and quantitative validity of Langevin model has been confirmed in the actual application to nuclear engineering field~(for example, see \cite{11randrup,15pahla,14aritomo,20sadhukhan}). 
In particular it is remarkable that fission data such as fission fragment yield (FFY) and total kinetic energy (TKE) are well reproduced by the Langevin calculations \cite{16usang} for isotopes with sufficient experimental data.
Although the Langevin model takes into account the stochastic aspect of fission at the level of macroscopic nucleus degree of freedom, the Langevin parameters are usually determined based on phenomenological treatment.
Since the Langevin model is represented by a kind of Newtonian equation of motion, quantum effects should be manually introduced by adjusting the parameter values.
On the other hand, the nuclear TDDFT takes into account the microscopic nucleon degrees of freedom without parameter fitting in terms of the reaction dynamics. 
In particular the TDDFT is a quantum theory in which quantum wave functions are directory calculated without any approximation.
Mainly due to the lack of rigorous treatment of stochastic process, the fission dynamics calculated by the TDDFT  (including the Time-Dependent Hartree-Fock Bogoliubov theory) is still necessary to be improved for reproducing observed fission phenomena (for the recent progress in such directions, see \cite{16bulgac,20bulgac-1,20bulgac-2,14wakhle,14obe,14sim,16ste,17tani}).
To compensate for the disadvantages of both the Langevin and the TDDFT models, it is practical to introduce a combined model incorporating stochastic and microscopic aspects, where there has not been any successful theoretical models including the stochastic aspect within the microscopic time-dependent calculations.
The combination of these two different models has been recently proposed by \cite{19sekizawa_hagino} and applied to three-dimensional Langevin calculations.
Here we propose a new type of TDDFT + Langevin model and applied to four dimensional Langevin calculations. 
Much attention is paid to revise the friction coefficients based on a microscopic and less phenomenological treatment of quantum dissipation.

In this paper, we systematically present microscopically calculated friction coefficients standing for the one-body dissipation, where the terminology ''one-body dissipation" means that the dissipation arising from the interaction between each nucleon and the summed-up potential field (cf. mean field).
The validity of the friction coefficients is confirmed by comparing to the experimental FFY.
Although conventional viscosity friction has already been treated in quasi fission dynamics, the appearance of inertial friction in nuclear fission processes is reported in this paper for the first time.  
As a result an energy-dependent transition of dissipation mechanism is quantitatively shown.

\section{TDDFT+Langevin model}
\subsection{Outline}
We study quasi-fission process by the four-dimensional Langevin model using the TDDFT-based friction force (TDDFT + Langevin model, for short). 
The main Langevin-part of the proposed model reads 
\begin{align}
\label{eq.langevin}
&\frac{dq_{\mu}}{dt} = m_{{\mu}j}^{-1}p_{j}, \\
&\frac{dp_{\mu}}{dt} = - \frac{dV}{dq_{i}} - {\frac{1}{2}}{\frac{d{m_{ij}^{-1}}}{dq_{\mu}}}p_{i}p_{j} - {\gamma}_{{\mu}i}m_{ij}^{-1}p_{j} - g_{{\mu}i}R_{i},
\end{align}
where three coefficients are the friction tensor $\gamma_{{\mu}i}$, the mass tensor $m_{ij}$ and the strength factor $g_{{\mu}i}$.
The $R_{j}$ is the white noise.
The function $V$ means nucleus-nucleus potential.
The unknown function $q_{\mu}$ is the Langevin coordinates, and $p_{\mu}$ means the momentum  ($\mu = 1, 2, 3$ and 4 in case of four dimensional calculation).
The friction tensor $\gamma_{{\mu}i}$ is usually obtained by the traditional Wall-and-Window formula~\cite{11randrup, 92abe}.
The Wall-and-Window formula is a well-known phenomenological formula in which a nucleus is regarded as a hydro-dynamical medium without holding any internal structure.
Note that there are several friction coefficients even based on some microscopic treatments (e.g., see \cite{97adamian} for a historical work, and see \cite{16usang} for a recent work).
In the proposed model, by generalizing the friction tensor, the following features are expected:
\begin{itemize}
\item energy dependence of the friction force,
\item isotope dependence of the friction force,
\item non-empirical quantum effect (cf. shell effect).
\end{itemize}
We take two steps for performing fission calculations.
First, by utilizing TDDFT calculations, the TDDFT trajectory $R(t)$ is obtained by measuring the relative distance between the colliding nuclei.
The energy-dependent trajectories lead to the energy-dependent component of friction tensor.
Second, by utilizing four-dimensional Langevin calculations with the TDDFT friction coefficients, the fission events are calculated.
By these two steps, both stochastic and microscopic aspects of fission dynamics are incorporated.

\subsection{Step 1: TDDFT part}
The numerical code Sky3D~\cite{14maruhn} is employed for the TDDFT calculations.
Space discretizations $dx, dy, dz$ are set to 1 fm and time step $dt$ to 0.2 fm/c in TDDFT part, and SV-bas effective nuclear interaction \cite{09klupfel} is employed.
The SV-bas interaction is especially known for reproducing the neutron skin thickness of heavy nuclei such as $^{208}$Pb (for a compilation of experimental and theoretical results, see \cite{15cosel,10reinhard,13reinhard}).
The skin thickness is expected to be relevant to the stability against the fission. 
For a mass-dependent comparison of binding energies between SV-bas calculations and experiments, see Table I of Ref. \cite{19iwata}, and for an individual role of each term, see Table I of Ref. \cite{15iwata}.
The benchmark results using different Skyrme parameters other than SV-bas are shown in Ref.~\cite{19nishikawa-3}.
For reference, preliminary calculations have been done for Sn isotopes (lighter cases) and $Z=120$ superheavy nuclei  (heavier cases) \cite{19nishikawa-1,19nishikawa-2}.

We carry out the TDDFT calculations of the early stage of nuclear reactions in symmetric central collisions
\begin{align}
\label{eq.reac}
^{A}Z+^{A}Z~\to~^{2A}2Z~\to~^{A}Z+^{A}Z
\end{align}
by focusing on the fission process ``$~^{2A}2Z~\to~^{A}Z+^{A}Z$ '', where $A$ and $Z$ are a mass number and a proton number, respectively.
The initial relative velocity of the collision ${\dot R}(0)$ is given, and the TDDFT wave function $\Psi(\bm{r},t)$ is obtained.
This calculation is used only for extracting the dissipation effect for a given nucleus $~^{2A}2Z$ and a given collision energy.
We calculate 25 heavy nuclei chosen from Uranium, Plutonium, Curium, Californium and Fermium isotopes.
The central (impact parameter = 0.0 fm) and symmetric (collision of identical nuclei) settings are advantageous in which the macroscopic friction coefficient at a fixed energy is determined referring to one collision event (for the method requiring plural number of collision events to determine one friction coefficient, see Ref.~\cite{08washiyama,09washiyama}).
This advantage contributes to obtain reliable energy-dependence of friction coefficients.

\begin{figure}[tb]
  \begin{center} 
    \includegraphics[width=9cm]{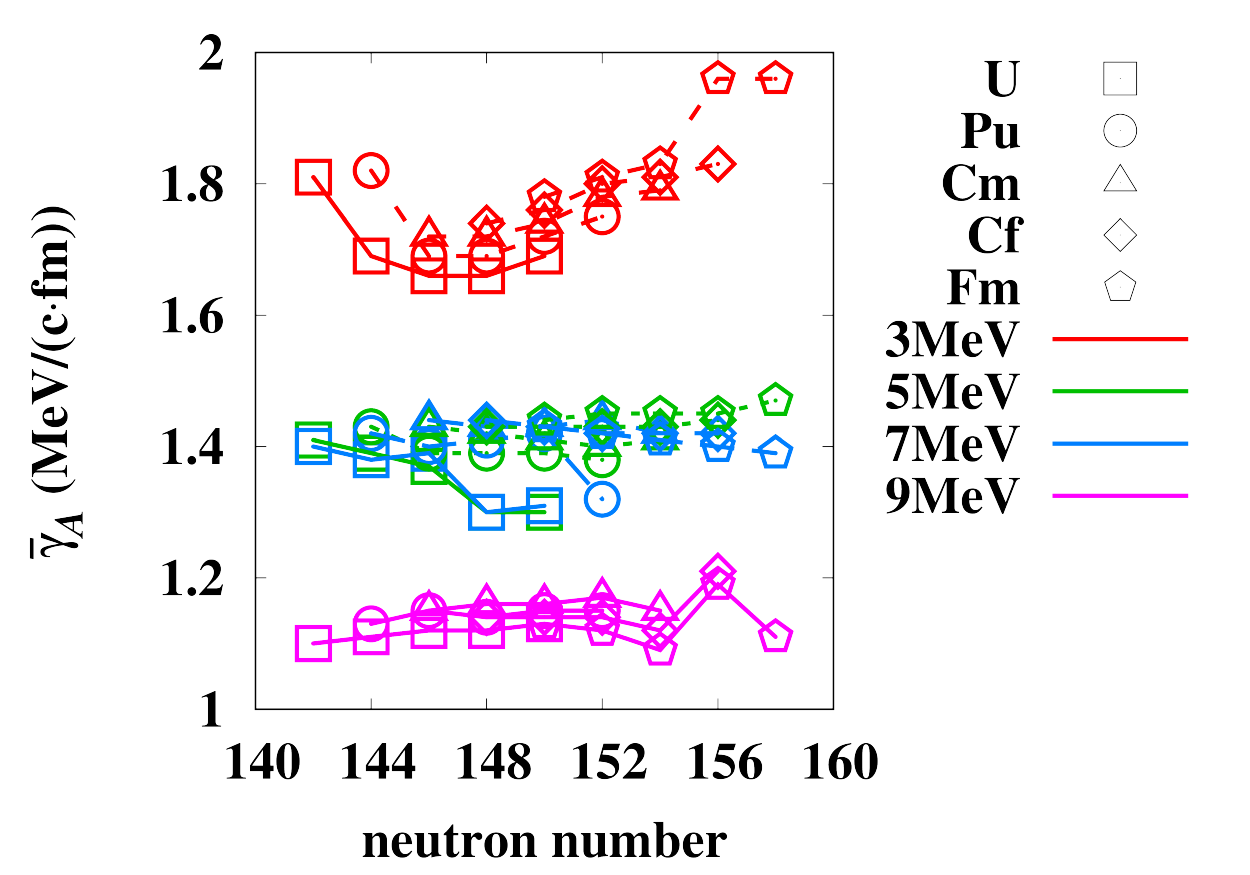}
    \caption{
(Color online) Energy and isotope dependence of the friction coefficients ($\bar{\gamma}_{A}$ = $\bar{\gamma}/2A$) for 25 nuclei. 
The difference of the chemical element is distinguished by the shape of graph legends.
The friction coefficients are localized to four different parts (four colors) depending on the collision energies $E/2A$ of 3.0, 5.0, 7.0 and 9.0~MeV. 
} 
    \label{fig.fc_high}
  \end{center}
\end{figure}

\begin{figure*}[t]
  \begin{center} 
    \includegraphics[width=14cm]{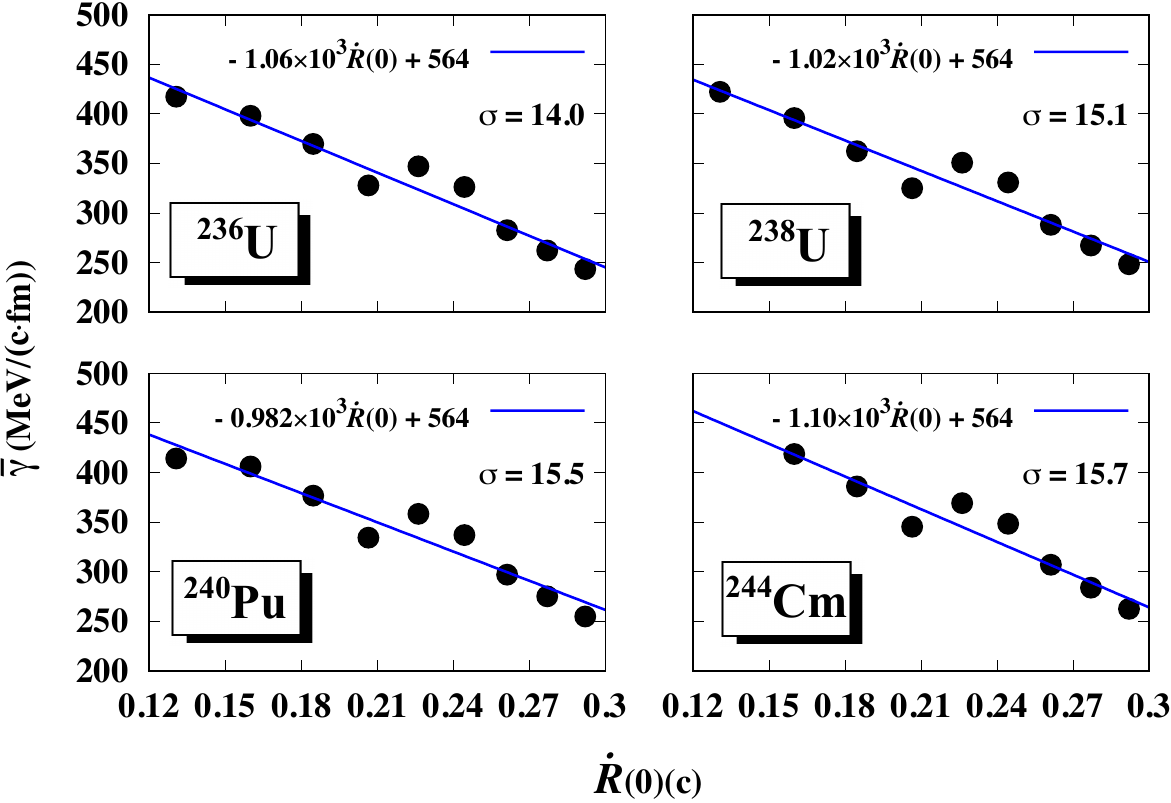}
  \end{center}
\caption{(Color online) Velocity-dependent friction coefficients for $^{236}$U, $^{238}$U, $^{240}$Pu and $^{244}$Cm.
The velocity is measured by its ratio to the speed of light $c$. 
Filled circles show the calculated values. 
The blue line is depicted by the linear regression.
The standard deviation of $\bar{\gamma}$ value compared to the depicted line is shown by $\sigma$ (MeV/fm$\cdot$c).}
\label{fig.extrap}
\end{figure*}

Let us move on to the relative distance $R(t)$ between the center-of-mass of two colliding nuclei from the TDDFT wave function.
Provided the TDDFT wave function $\Psi(\bm{r},t)$ depending on space $\bm{r} = (x, y, z)$ and time $t$, the TDDFT trajectory $R(t)$ is obtained as a collective quantity
\begin{align}
\label{eq.tra}
R(t)={\iiint}2|z||\Psi(\bm{r},t)|^{2}dxdydz,
\end{align}
where $z$ denotes the collision axis, and the initial positions of colliding nuclei are located on $z$-axis to be symmetric with respect to the $z = 0$ plane.
The TDDFT trajectory $R(t)$, which is calculated based on the TDDFT wave function $\Psi(\bm{r},t)$, includes the one-body quantum dissipation effect without any loss.
Although it is not explicitly shown in the representation, the trajectory $R(t)$ is obtained depending also on the initial relative velocity $\dot{R}(0)$ and therefore the collision energy per nucleon
\begin{equation} \label{eovera}
E/2A = \frac{1}{2} m \left( \dot{R}(0) \right)^2,
\end{equation}
where $m=$938~MeV is the nucleon mass.
The detailed definition of energy-dependent trajectory $R(t)$ is explained in Ref.~\cite{11iwata}. 
Following the preceding works \cite{08washiyama,09washiyama,11iwata}, we assume that $R(t)$ satisfies the Newtonian equation of motion
\begin{align}
\label{eq.ce} 
{\mu}{\ddot{R}}(t)+\frac{dV(E)}{dR}+ \overline{\gamma}(E){\dot{R}}(t) = 0 \ ,
\end{align}
where $\ddot{R}(t)$ and $\dot{R}(t)$ are prepared by differentiating $R(t)$, and the initial distance $R(0)$ and the initial velocity $\dot{R}(0)$ are given by the initial setting. 
The coefficients $ \overline{\gamma}$  and $V$ denote the nucleus-nucleus friction coefficient and the nucleus-nucleus potential, respectively. 
Since Eq.~\eqref{eq.ce} holds the same form as the Langevin equation, this treatment provides an interface between the TDDFT and the Langevin dynamics.
Finally, the energy-dependent friction coefficient is obtained by integrating Eq.~\eqref{eq.ce}
\begin{align}
\label{eq.barg}
\overline{\gamma}(E) = \frac{\{{\frac{1}{2}}\mu{\dot{R}}^2(t_{i}) + V(t_{i})\}-\{{\frac{1}{2}}\mu{\dot{R}}^2(t_{f}) + V(t_{f})\}}{\int_{t_{i}}^{t_{f}}\{\dot{R}(t)\}^{2}dt} \
\end{align}
without introducing any further approximation, where $t_i$ and $t_f$ denote the initial time and final time, respectively.
This is a kind of inverse problem in which the unknown parameters are determined by the known solution $R(t)$ of Eq.~(\ref{eq.ce}).
In this way the quantum one-body dissipation, which has been suggested to be associated with the spin-orbit component of the nuclear force \cite{12iwata} based on the TDDFT calculations, is made into the energy-dependent friction coefficient (to a lesser degree, one-body dissipation has been suggested to be associated with the spin-current tensor part of the nuclear force \cite{11iwata, 18guo}).

As a result the energy-dependent friction coefficient is calculated for each excited nucleus $^{2A}2Z$.
Comparing to the conventional Langevin calculation using the Wall-and-window formula type energy-independent friction coefficient, it is remarkable that the energy dependence is introduced in a non-empirical manner.

\subsection{Energetic extrapolation}
In order to find the friction coefficients at the extra energies, which are too low to be treated by the TDDFT, the extrapolation method is introduced. 
The standard energy range of the TDDFT calculations (mean-field energy, for short) is located from the Coulomb barrier energy (several times of 100 keV per nucleon) to the almost twice of the Fermi energy (several times of 10 MeV per nucleon) \cite{10iwata}.
Since fission appears even below the Coulomb barrier energy, this treatment is necessary to calculate some realistic fission events. 
That is, the energetic extrapolation enables us to access the dynamics at beyond the mean-field energies.

\begin{table}[tb] 
\caption{Table for slope $a$ and $b$ for 25 heavy compound nuclei based on the TDDFT friction coefficients, which can be used for most of fission events with the excitation energy less than a few MeV per nucleon. The value given in parenthesis is the uncertainty.
The uncertainty is statistically estimated by the standard deviation.}
\begin{tabular}{c|c|c}
Nucleus & $a$$(\rm {1/(fm{\cdot}c^2)})$ & $b$$(\rm {MeV/(fm{\cdot}c)})$ \\ \hline
$^{234}$U  & -1217($\pm$123) & 608($\pm$27) \\
$^{236}$U  & -1063($\pm$92)  & 565($\pm$19) \\
$^{238}$U  & -1022($\pm$86)  & 557($\pm$17) \\
$^{240}$U  & -921($\pm$109)  & 529($\pm$27) \\
$^{242}$U  & -1042($\pm$124) & 562($\pm$33) \\
$^{236}$Pu & -1181($\pm$146) & 609($\pm$32) \\
$^{238}$Pu & -983($\pm$119)  & 556($\pm$26) \\
$^{240}$Pu & -985($\pm$113)  & 556($\pm$25) \\
$^{242}$Pu & -1101($\pm$133) & 590($\pm$33) \\
$^{244}$Pu & -1175($\pm$126) & 608($\pm$34) \\
$^{238}$Cm & -1156($\pm$120) & 606($\pm$28) \\
$^{240}$Cm & -1097($\pm$120) & 594($\pm$28) \\
$^{242}$Cm & -1138($\pm$122) & 607($\pm$30) \\
$^{244}$Cm & -1161($\pm$144) & 617($\pm$37) \\
$^{246}$Cm & -1202($\pm$139) & 631($\pm$36) \\
$^{240}$Cf & -1202($\pm$129) & 623($\pm$30) \\
$^{242}$Cf & -1236($\pm$118) & 634($\pm$29) \\
$^{244}$Cf & -1176($\pm$127) & 622($\pm$30) \\
$^{246}$Cf & -1326($\pm$126) & 661($\pm$32) \\
$^{248}$Cf & -1329($\pm$144) & 666($\pm$36) \\
$^{242}$Fm & -1259($\pm$98)  & 642($\pm$21) \\
$^{244}$Fm & -1391($\pm$106) & 677($\pm$27) \\
$^{246}$Fm & -1463($\pm$120) & 695($\pm$29) \\
$^{248}$Fm & -1541($\pm$218) & 723($\pm$57) \\
$^{250}$Fm & -1616($\pm$181) & 744($\pm$47) \\
\hline \hline
\end{tabular}
\label{t_slope}
\end{table}

\begin{figure*}[t]
  \begin{center} 
    \hspace{-6mm}
    \includegraphics[width=12cm]{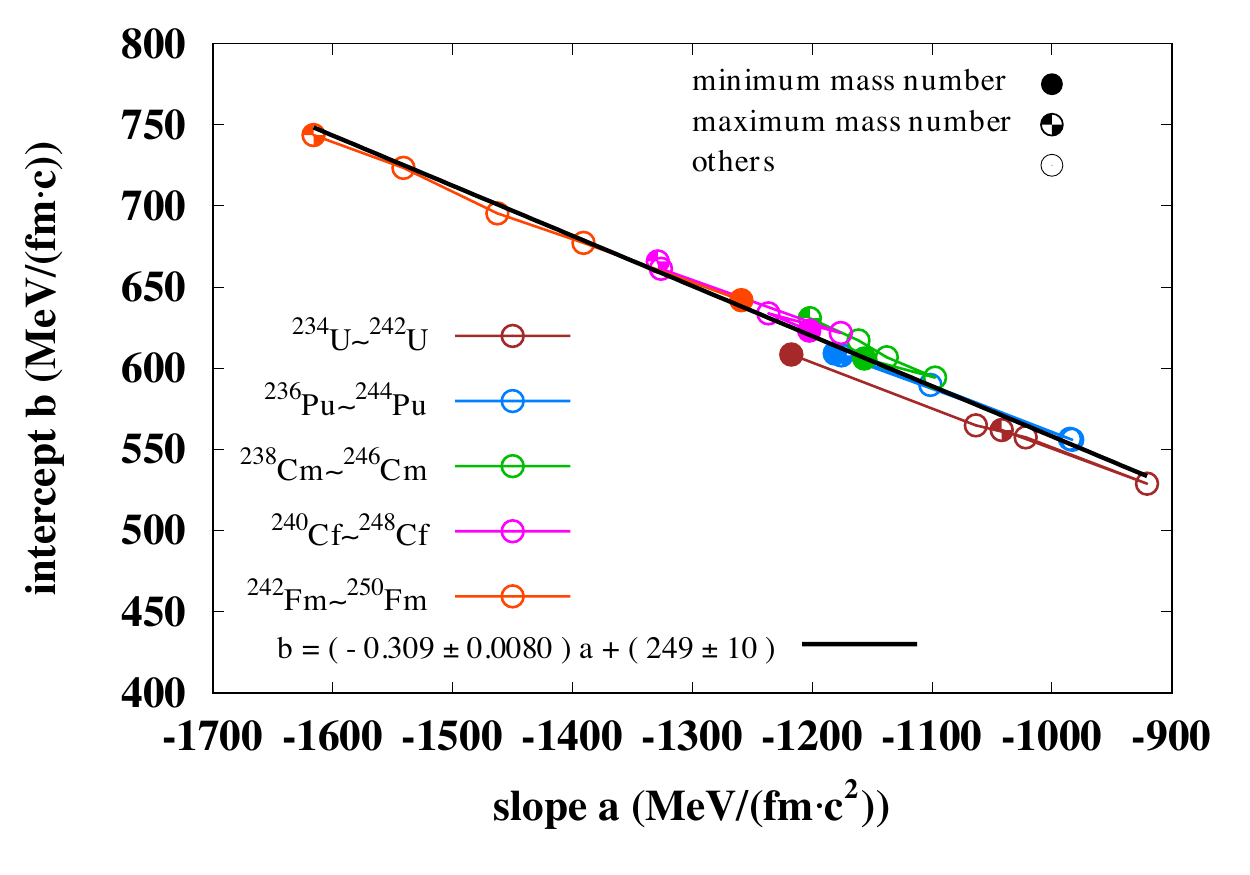}
    \caption{ (Color online) The relation between the coefficients $a$ and $b$.
      Slope $a$ and intercept $b$ of Eq.~\eqref{eq-reg} are shown for the 25 nuclei, where the values for the same chemical element are connected for guiding eyes (isotope-chain).
The linear regression is implemented, and the obtained line is shown by a thick line.
The standard deviation is adopted for estimating errors, where the proportional ratio $db/da=-0.309$ possibly includes the error $\pm 0.0080$ (2.5$\%$) and  the reference value $b(0)=249$ includes the error $\pm 10$ (4.0$\%$). } 
  \label{fig.aandb}
  \end{center}
  \begin{center} 
    \includegraphics[width=8cm]{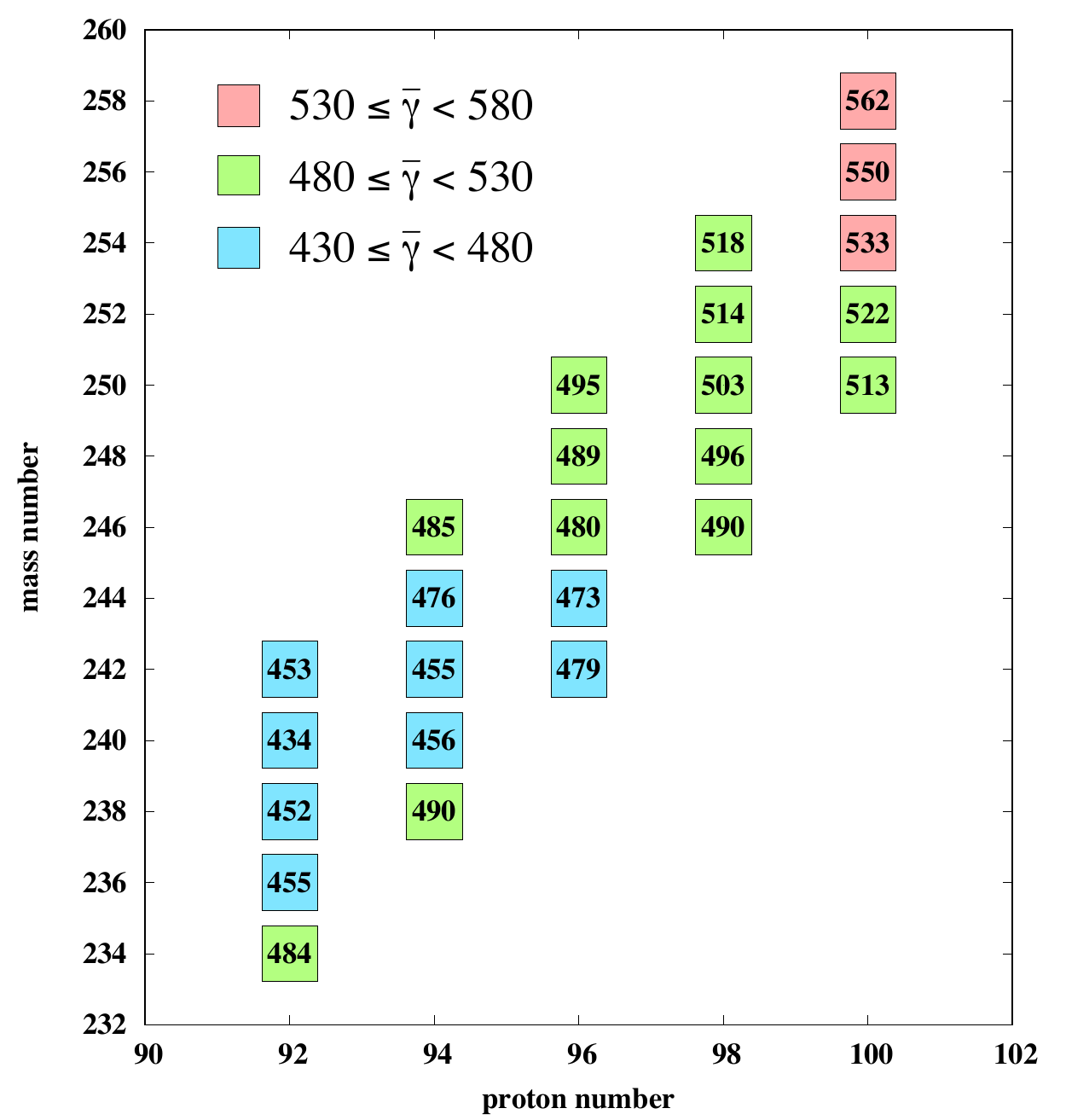}
  \end{center}
 \caption{(Color online) Landscape of the friction coefficients for U, Pu, Cm, Cf, and Fm isotopes.
 The corresponding energy (20 MeV as a total energy the laboratory frame) is small enough to be below the Coulomb barrier energy.} 
 \label{fig.scope}
\end{figure*}

For the mean-field energies, we calculate friction coefficients for the collision energies $E/2A =$  2, 3, 4,$ \cdots,$ 10~MeV (cf. Eq.~(\ref{eovera})).
Energy and isotope dependence of friction coefficients is shown in Fig.~\ref{fig.fc_high}.
As a general trend in the mean-field energies, the friction coefficients becomes larger for lower energy collisions.
Furthermore, as a common feature, the isotope dependence of the friction coefficient is more complicated at lower energies and monotonous at higher energies.
Indeed, in case of $E/2A = 9$ MeV, the friction coefficient is almost constant equal to 1.10 MeV/(c$\cdot$fm).
It simply shows the complexity of the nonlinear dynamics for lower energy cases.

Figure~{\ref{fig.extrap}} demonstrates the velocity dependence of the friction coefficient (cf. Eq.~\eqref{eovera}).  
As shown in Fig.~{\ref{fig.extrap}}, friction coefficients are well approximated by a linear segment, and extrapolation is performed by the linear regression in which the two coefficients $a$ and $b$ of
\begin{equation} \begin{array}{ll} \label{eq-reg}
\overline{\gamma}(E)  = a \left( \sqrt{2E/{\mu}} \right) + b \vspace{1.5mm} \\
\qquad  =  a{\dot{R}}(0) + b
\end{array} \end{equation}
are determined by the least square method.
Each value of $a$ and $b$ can be different depending on the isotope. 
The calculated results for slope $a$ and the intercept $b$ for 25 heavy nuclei are listed in Table~{\ref{t_slope}}, where the uncertainly of $a$ and $b$ are also shown.
The good agreement with the linearity provides us a sound motivation to introduce an energetic extrapolation particularly for lower energies in which most of fission events take place.

The relation between the parameters $a$ and $b$ is shown in  Fig.~{\ref{fig.aandb}}.
The $(a,b)$ values of 25 nuclei do not distribute in an entire area defined by $-1615 \le a \le -920$ and $528 \le b \le 744$.
More precisely a simple linear dependence is found as
\begin{equation}
b = - 0.309a + 249
\end{equation}
being independent of the difference of the chemical elements.
Consequently the energy-dependent friction force is represented using a single parameter as
\begin{equation} \label{eq-reg-onlya}
\overline{\gamma}(E)  = a {\dot{R}}(0) + ( - 0.309a + 249 ).
\end{equation}
Eq.~\eqref{eq-reg-onlya} means that the parameter is determined by choosing a compound nucleus  (cf. the amplitude of $a$ in Table \ref{t_slope}).

\begin{figure*}[tb]
  \begin{center} 
    \includegraphics[width=14cm]{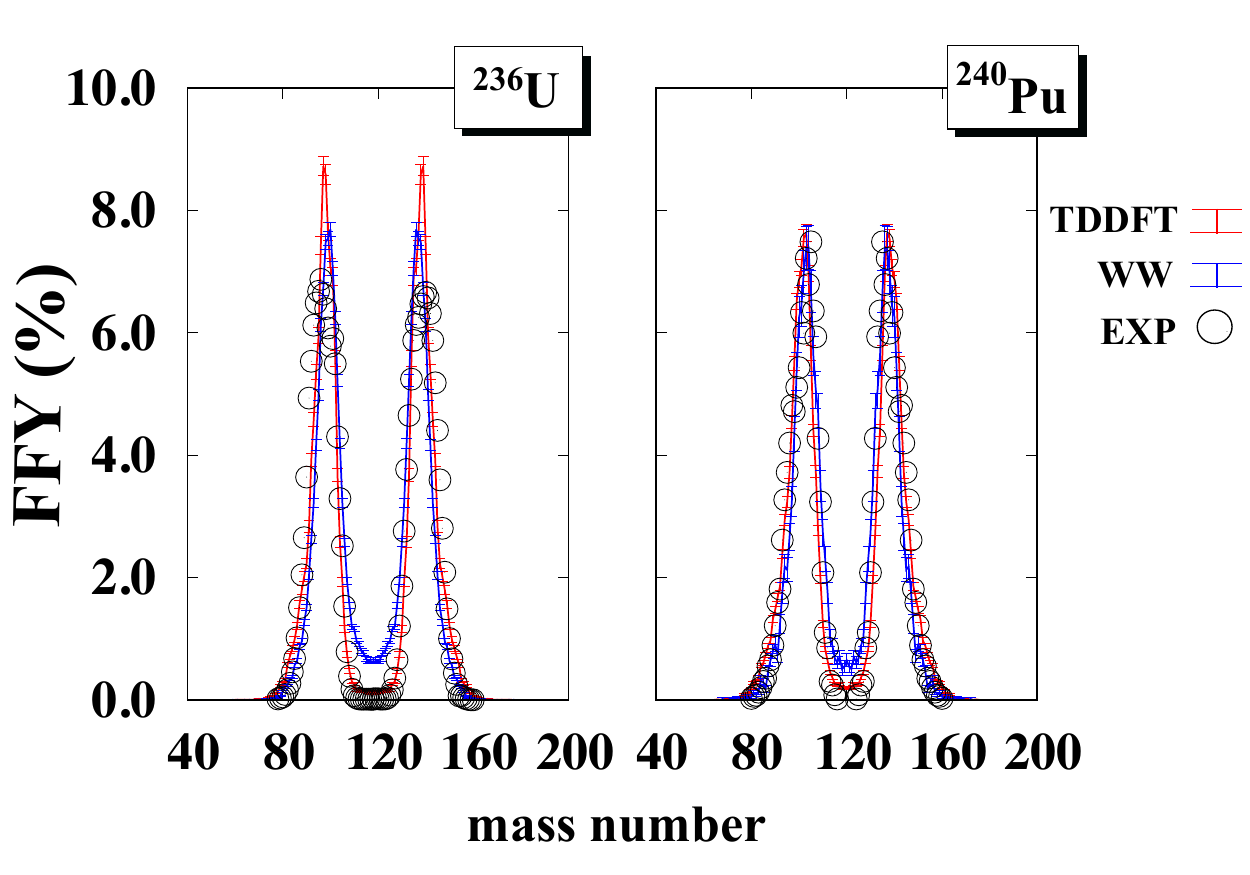}
  \end{center}
    \caption{
(Color online) FFY of $^{236}$U and $^{240}$Pu fission with the excitation energy $E^* =$ 20 MeV. 
The FFY calculated by the TDDFT+Langevin method is shown by red line and the FFY  by the conventional Langevin calculation simply using the Wall-and-Window formula is shown by blue line.
Open circles are the experimental data (Refs.~\cite{87straede, 06zenyalov} for $^{236}$U, and Ref.~\cite{97dematt} for $^{240}$Pu).
\label{fig.ffy}}
\end{figure*}

\subsection{Step 2: Langevin part}
The fission dynamics of $^{2A}2Z$ is calculated by recently developed 4D-Langevin code~\cite{16usang,17ishizuka}.
Time step of the Langevin calculation is fixed to 0.1 fm/c.
We collect 100,000 events to have one Langevin statistics.
Let us introduce $\overline{\gamma}(E)$ to the Langevin calculations.
The friction coefficient has the tensor form $\bm{\gamma}=\{\gamma_{ij}\}$ with $1\leq {i,j} \leq 4$.
The friction tensor $\bm{\gamma}$ in the proposed model is calculated by
\begin{equation} \begin{array}{ll}
&\gamma_{{\mu}i} (E) = \overline{\gamma}(E) \quad ({\mu}=i=1), \vspace{2.5mm} \\
&\gamma_{{\mu}i}=g_{{\mu}i}{\quad}{\quad} (\rm{others}),
\end{array} \end{equation}
where $g_{ij}$ denotes the energy-independent friction coefficient calculated by the Wall-and-Window formula, and $\gamma_{11}$ is replaced with the energy-dependent friction coefficient $\overline{\gamma}(E)$.
The friction $\gamma_{11}$ is acting on the motion of two center-of-masses, and $\gamma_{11}$ is expected to be rather highly energy-dependent compared to the other components.
Indeed the friction components other than  $\gamma_{11}$ acting on changing nuclear shape are indirect effects in terms of relaxing the center-of-mass motion.

 \begin{figure*}[t]
 \begin{center}
    \includegraphics[width=8.5cm]{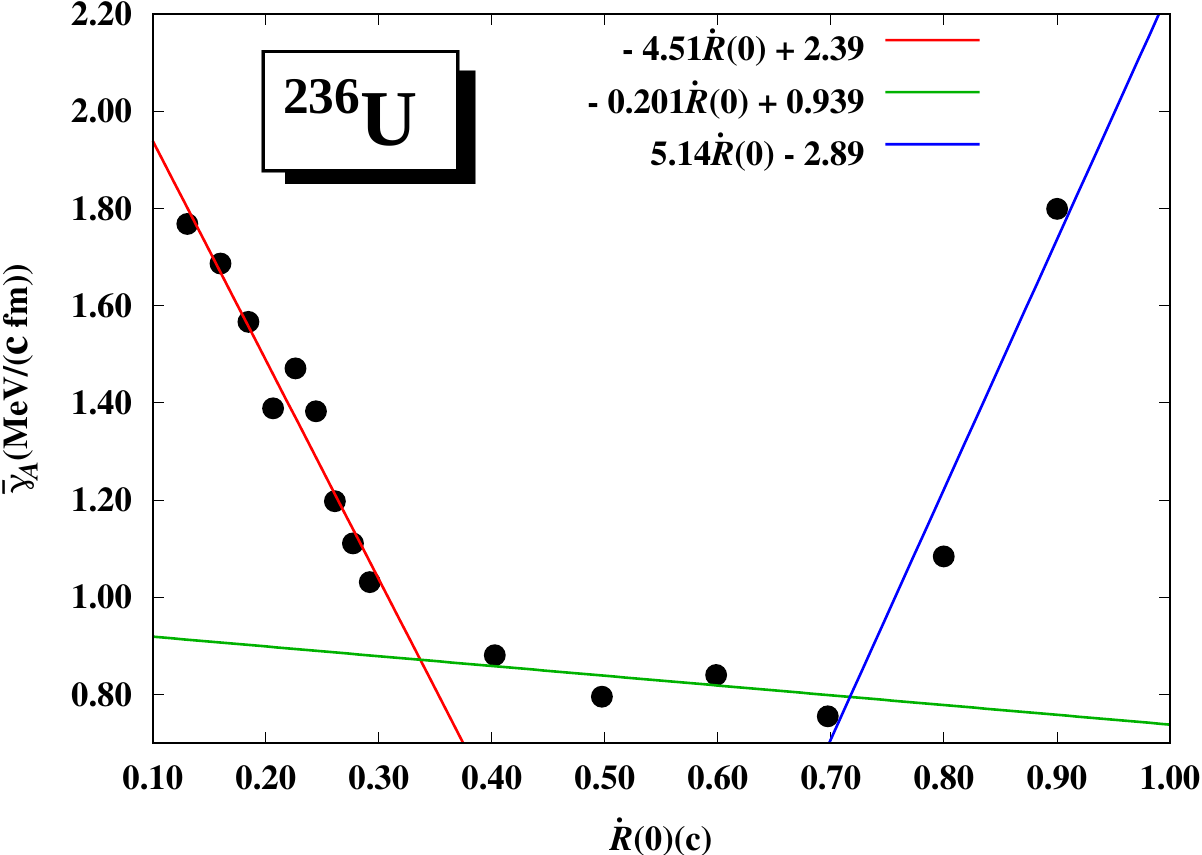} \quad
    \includegraphics[width=8.5cm]{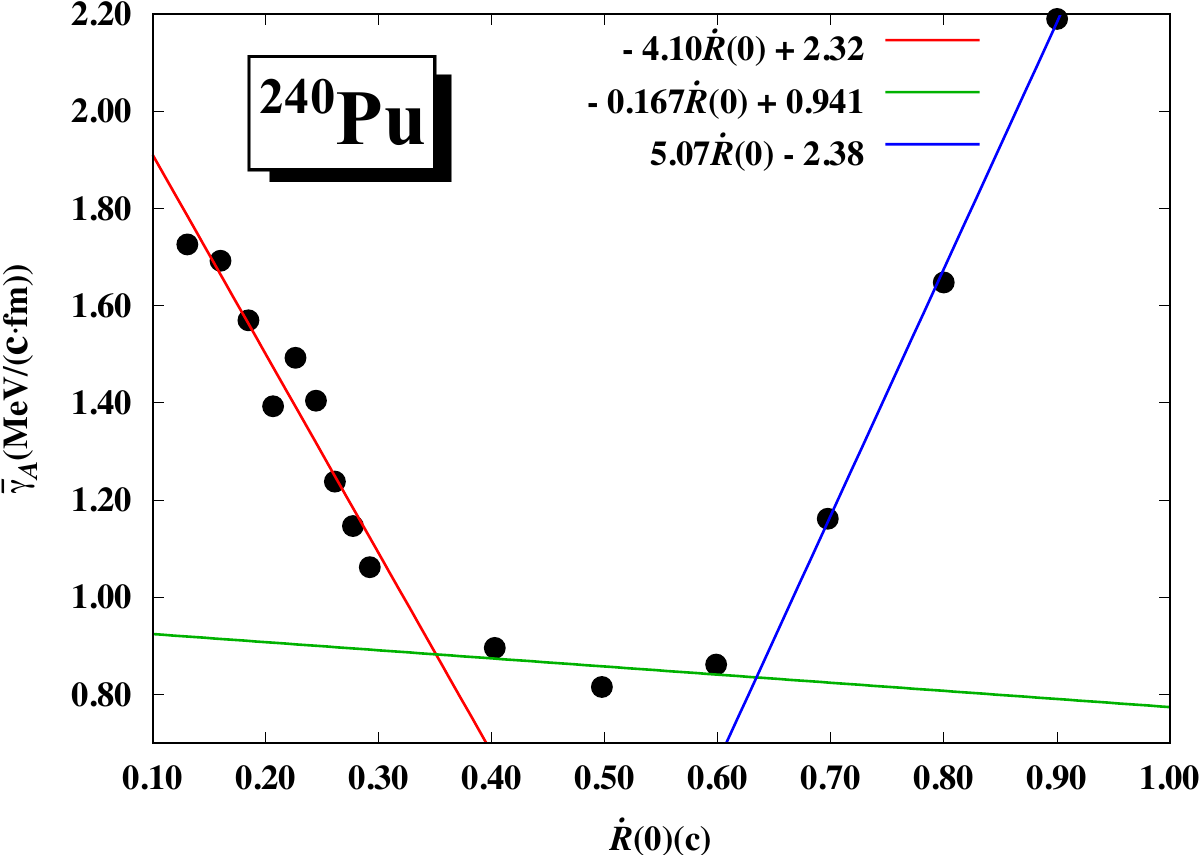}
 \end{center}
 \caption{(Color online) Velocity dependence of the friction coefficient  $\bar{\gamma}_{A}$ = $\bar{\gamma}/2A$ derived from the  TDDFT trajectory.
The friction coefficients for $^{236}$U and for $^{240}$Pu are shown with the linear regression lines in left panel and right panels respectively.
The red line corresponds to the friction coefficients at lower energies ($E/2A \le 10$ MeV), the green lines to those at intermediate energies, and blue lines to those at higher energy ($E/2A > 30$ MeV).
The crossing points, which mean the transition energy of dissipation mechanism, are located at 0.34$c$ and 0.72$c$ in case of $^{236}$U, and  at 0.35$c$ and 0.63$c$ in case of $^{240}$Pu. }
 \label{fig.fricoeffs}
\end{figure*}

In order to apply the friction shown in Eq.~\eqref{eq-reg-onlya} to the Langevin calculation, the total excitation energy $E^*$ is necessary.
The total excitation energy of compound nucleus 
\begin{equation}
E^* =  E + Q
\end{equation} 
is usually an input of the Langevin calculation, where $Q$ is the Q-value of the fission reaction and $E$ is the collision energy.
Although the collision energy per nucleon $E/2A$ of  TDDFT calculations is measured in the center-of-mass frame, the excitation energy $E^*$ is usually represented in the laboratory frame.
In this situations it is necessary to translate center-of-mass energy into the energy in the laboratory frame  (for the translation, we use the equations shown in \cite{13iwata-4}).

\section{Result}
\subsection{Fission fragment yields}
\label{vopt}
Let us begin with the cases with its excitation energy $E^* =20$ MeV.
Since fission events at the 20 MeV ($\sim$ 100 keV per nucleon) excitation energy is out of the TDDFT description, the extrapolation method is utilized.
For the excitation energy $E^* =20$ MeV, the friction coefficient $\gamma_{11}(E)$ are 452 MeV/(fm$\cdot$c) and 448 MeV/(fm$\cdot$c) for $^{236}$U and $^{240}$Pu, respectively.
Figure~\ref{fig.scope} summarizes the friction coefficients at the energy below the Coulomb barrier energy.
Even in a fixed energy, we see that the $\overline{\gamma}$ depends both on mass number and proton number.
As a general trend, $\overline{\gamma}$ values become larger for larger mass and proton numbers.

For the validity of proposed method including the energetic extrapolation, we calculate FFYs by the TDDFT+Langevin model, and they are compared to both experiments and conventional Langevin calculations.
Low-energy cases beyond the mean-field energies are studied in Fig.~\ref{fig.ffy}.
Left panel of Fig.~\ref{fig.ffy} compares the calculated FFY of $^{236}$U to experimental data.
Two peaks located around mass numbers 100 and 140 are well reproduced.
In addition, symmetric fission component of the yield between 100 and 135 agrees to experiment better for the proposed calculation than for the conventional calculation.
Right panel of Fig.~\ref{fig.ffy} compares the calculated FFY of $^{240}$Pu to experimental data.
Also in case of $^{240}$Pu an improvement in the symmetric fission part is seen.
For the improvement compared to the conventional Langevin model, 0.64 $\%$ and 0.59 $\%$ excesses of the symmetric fission probabilities of $^{236}$U and $^{240}$Pu are improved to 0.05 $\%$ and 0.26 $\%$ respectively.  
According to these agreements of the FFYs, we see that $\gamma_{11}$ (the dissipation arising from the center-of-mass relative motion) plays a prominent role in the FFY distribution.
Similar improvement can be seen in some different cases other than $^{236}$U and $^{240}$Pu.
For agreement between the TDDFT+Langevin calculations using some different Skyrme interactions and experimental data can be found in our preliminary calculations~\cite{19nishikawa-2}.

\begin{figure*}[t]
  \begin{center} 
    \includegraphics[width=18cm]{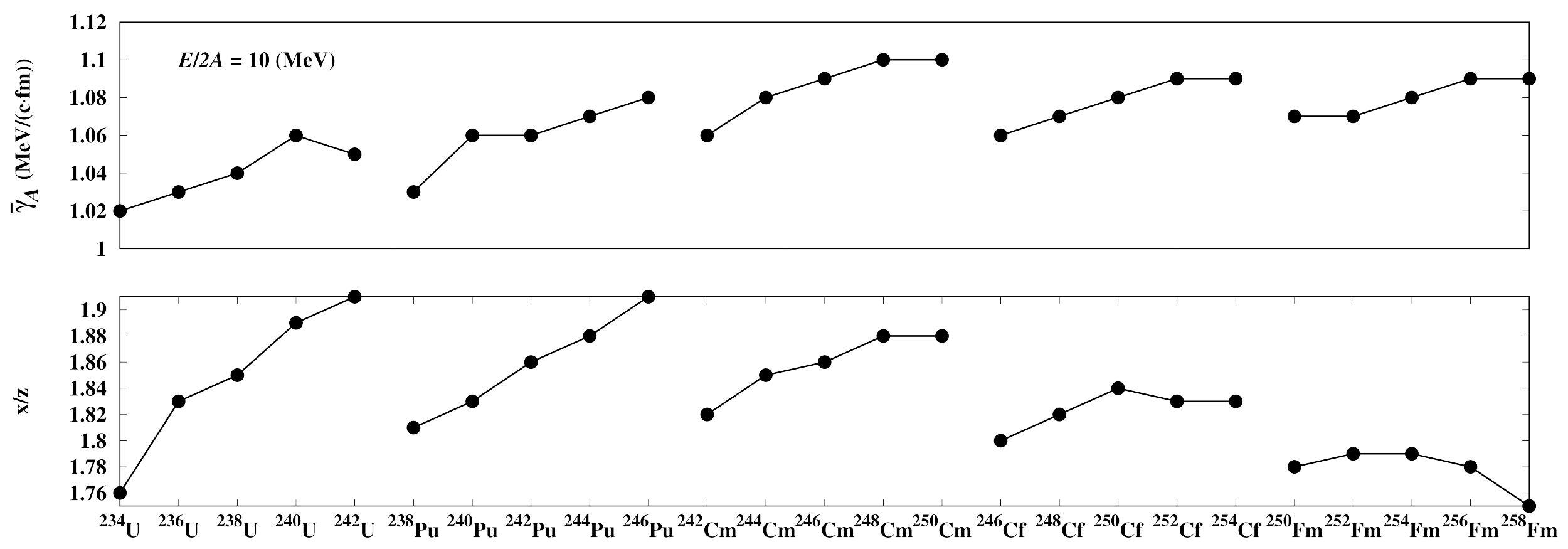}
    \caption{
(Color online) Deformations of ground states are compared to the friction coefficients $\bar{\gamma}_{A}$ = $\bar{\gamma}/2A$.
Here we consider a binary system consisting of  two $^{A}Z$,  if fission appears.
The deformation factor $x/z$ is calculated by the longest length $x$ and the shortest length $z$ of a fissioning binary system (consisting of  two $^{A}Z$).
The deformed ground state is obtained by the static DFT calculations employing the same Skyrme parameter set (SV-bas).
For achieving fission, a  nucleus  $^{A}Z$ experiences a friction force with its path length being roughly given by $x$.
Mediated by each path length $x$, the deformation is related to the total amount of dissipation during a fission process.
}
    \label{fig.shapeandg}
       \end{center}
\end{figure*}

\subsection{Appearance of inertial friction}
In the Langevin calculations, macroscopic friction acting not on nucleons but on the nuclei is used.
For the macroscopic friction effect, the friction of two kinds are known: viscous friction and inertial friction.
The viscous friction force is proportional to $\dot{R}(t)$, while the inertial friction force is proportional to~${\dot{R}}^2(t)$.
Qualitatively speaking, the inertial friction is caused by pushing the existing medium aside, so that the inertial friction is expected to appear in the collision with large relative velocity.
The appearance of both types of friction effect is suggested by Eq.~\eqref{eq-reg-onlya}. 
Indeed, Eqs~\eqref{eovera}, \eqref{eq-reg}, and ~\eqref{eq-reg-onlya} lead to the macroscopic friction force
\begin{equation} \begin{array}{ll} \label{eq.frictionformula}
 \gamma_{11}(E) {\dot{R}(t)}=  a {\dot{R}(t)}  {\dot{R}(0)} + b\dot{R}(t)  \vspace{1.5mm} \\
 \quad \sim  a {\dot{R}(t)}  {\dot{R}(0)} +  ( - 0.309a + 249 ) \dot{R}(t). 
\end{array} \end{equation}
Since the first term of the right hand side of Eq.~\eqref{eq.frictionformula} is proportional to ${\dot R}^2$, this term behaves as the inertial friction.
The second term, which are introduced in conventional calculations (for a textbook containing the historical development of nuclear reaction theory, see \cite{96froblich}), behaves as the viscous friction force.
Note that the coefficient $a$ in Eq.~\eqref{eq.frictionformula} is assumed to be zero in conventional Langevin calculations, and suggested to be nonzero in the present paper (cf Table I). 

Although  the inertial friction is found typically in the higher energies, such an effect are found in both lower and higher energy parts (Fig.~{\ref{fig.fricoeffs}}).
The friction types are classified into three parts: inertial friction with negative $a$ in lower energies, viscous friction with almost zero $a$ in intermediate energies, and inertial friction with positive $a$ in higher energies.
With respect to the fission analysis, we definitely use the coefficient with ${\dot R}(0) \le 0.30$~c.
The high energy cases, which are too high to be treated by the TDDFT, are shown for theoretical interest to clarify the detailed features of one-body quantum dissipation.
In fact, for the realistic treatment of higher energy collisions it is necessary to introduce the dissipation/friction arising from the nucleon-nucleon collision effect, where the TDDFT is a collisionless quantum framework in terms of nucleon-nucleon collisions.

\subsubsection{Inertial friction at lower energies} 
The low-energy heavy-ion collisions with those relative velocities less than 1/3 of speed of light are nonlinear
processes (Fig.~\ref{fig.fc_high}).
In low energy part with ${\dot R(0)} \le 0.30~c$, the friction becomes smaller depending on the energy increase.
The mean-field energy, which is mentioned in the previous section, is located around this low energy part.
The energy dependence can be roughly understood by the duration time being larger for lower energies.
Here is the fact that energy dissipation is expected to be larger if the touching duration time becomes larger.
The same explanation can be roughly valid to the mass dependence that the friction becomes larger for a nucleus with heavier mass.
The concept of duration time essentially arises from the finite-body property of nucleus-nucleus collisions, so that the concept of duration time does not directly show up in the infinite systems (e.g., fluid or hydro-dynamical models), and the concept of mean free path is essential in infinite systems, instead. Here is a reason why the inertial energy dissipation with negative coefficient is initially or uniquely detected in nucleus-nucleus collisions.
Meanwhile a key to understand the details of coefficients with negative $a$ in low energy events is the shape of nucleus.
 It allows us to access the nonlinearity of low-energy collisions.
As shown in  Fig.~\ref{fig.shapeandg}, the isotope dependence of $\bar{\gamma}$ is found out to be roughly approximated by that of deformation factor.
Since the inertial friction is caused by pushing medium aside, the friction caused by the deformation is responsible for the inertial friction in very low energy collisions to a non-negligible extent. 
Consequently the inertial friction at lower energies can be understood by the finite-body property and the nonlinearity in a mass- and momentum-dependent manner.

 \subsubsection{Viscous friction at intermediate energies}

In intermediate energy part with $0.35~c < {\dot R(0)} < 0.60~c$, the friction becomes very small and almost constant being independent of the energy.
In this energy part, we found that the inertial friction is smaller compared to the viscous friction; indeed the amplitude of viscous friction coefficient  $\bar{\gamma_A}$ is roughly equal to 1.00 MeV/(c$\cdot$fm) for both $^{236}$U and $^{240}$Pu, while the amplitude of inertial friction is estimated to be around $0.20 \times 0.50 = 0.100$ MeV/(c$\cdot$fm) for $^{236}$ and $0.170 \times 0.50 =0.085$ MeV/(c$\cdot$ fm) for $^{240}$Pu (Fig.~\ref{fig.fricoeffs}).
The conventional energy independent friction coefficient is applicable to this intermediate energy part with $E/2A \sim 20 $ to 30 MeV.

 \subsubsection{Inertial friction at higher energies}
In high energy part with $0.70~c \le {\dot R(0)}$, the friction becomes larger depending on the energy increase.
That is, the inertial dissipation depending on $\dot{R}^2$ appears.
Note that the sign of friction coefficient $a$ is opposite compared to the low energy case.
This is the inertial dissipation in the ordinary sense.
Since the collision energy is above the mean-field energy and also above the non-relativistic energy limit, we will show them only for the reference to see the energy-dependent behavior of one-body quantum dissipation.
Consequently, for all the systematic calculations, the TDDFT calculations show the inertial dissipation in the ordinary sense, if the relative velocity of the collision is larger than 70$\%$ of the speed of light.

\section{Conclusion} 
We proposed the TDDFT + four-dimensional Langevin model, and the energy-dependent and microscopic friction coefficients are  non-empirical ly introduced to the Langevin calculations.
The agreement between the calculated FFYs and the experimental results show a systematical improvement compared to the conventional Langevin calculations (Fig.~{\ref{fig.ffy}}).
The introduction of energy-dependent and non-empirical friction coefficient, which is equivalent to take the inertial friction into account, is expected to have an impact on the existing and future Langevin calculations.
The energy-dependent friction coefficient \eqref{eq.frictionformula} with its parameter value is available in Table I.

Based on the TDDFT + four-dimensional Langevin model, we study the details of the dissipation mechanism during nuclear fission.
Three different types of dissipation mechanism are noticed (Fig.~{\ref{fig.fricoeffs}});
inertial energy dissipation being associated with the nuclear deformation appears at low energy part (Fig.~{\ref{fig.shapeandg}}),
the dominance of viscous energy dissipation appears at intermediate energy part, and inertial energy dissipation of the ordinary type appears at high energy part.
Even if we limit ourselves to the one-body quantum dissipation, we see that three different types of dissipation mechanism appear depending on the energy. 
At the lower and higher energy part the appearance of inertial dissipation is directly associated with the appearance of nonlinearity.
At the intermediate energy part with its energy per nucleon roughly equal to 20 MeV (with its relative velocity roughly equal to 1/3 of the speed of light), the dissipation effect itself becomes quite small.
It means that the fissioning nucleus experiences dynamics similar to the transparency.
In particular, the appearance of inertial friction at lower energies shows a role of dissipation in the nuclear fission process.

Transition of dissipation mechanism is suggested to exist.
The lower energy bounds of intermediate energies correspond to the upper energy limit of the fast charge equilibration \cite{10iwata} in which the nuclear soliton (more precisely, nuclear non-topological soliton \cite{92lee}) is suggested to appear in reactions involving light nuclei \cite{15iwata,19iwata, 20iwata}.
Although soliton is claimed to exist in reactions between light nuclei with their total mass numbers satisfying $A \le 16$ \cite{19iwata, 20iwata}, the TDDFT + four-dimensional Langevin model captures the remaining trace of mass-dependent existence of nuclear solitons.
Each boundary of lower, intermediate and higher energy parts holds the physical meanings; the upper boundary is associated the transition of dominant wave propagation, and the lower boundary is associated with the transition of  dominant equilibration dynamics. 
The former transition-energy is common to the hydro dynamics, and the latter transition-energy is specific to nuclear dynamics which includes two different medium (protons and neutrons) and  the charge equilibration. 
The evaluation of dissipation effect due to the mass tensor, which can be studied by a simple extension of the present method without changing the TDDFT part, is a future problem.
 \vspace{6mm} \\


This work was partially supported by JSPS KAKENHI (Grant No. 17K05440). 
The numerical calculations are carried out at the Yukawa Institute Computer Facility, at the high-performance workstation system of Tokyo Institute of Technology and at workstation system at Kansai University.


\end{document}